\documentclass[12pt]{article}
\usepackage{graphicx}
\usepackage[hang,small,bf]{caption}
\newcommand{\be}[3]{\begin{equation}  \label{#1#2#3}}
\newcommand{\ee}{ \end{equation}}
\newcommand{\ba}{\begin{array}}
\newcommand{\ea}{\end{array}}
\begin {document}

%%%%%%%%%%%%%%%%%%%%%%%%%%%%%%%%%%%%%%%%%%%%%%%%%%%%%%%%%%%%%%%%%%%

\thispagestyle{empty}
\rightline{HU-EP-00/64}
\rightline{hep-th/0101212}

\vspace{18truemm}

\centerline{\bf \Large 
Domain walls and flow equations in supergravity
}

\bigskip

\vspace{2truecm}

\centerline{\bf Klaus Behrndt\footnote{e-mail: 
 behrndt@physik.hu-berlin.de} }

\vspace{1truecm}
\centerline{\em Institut f\"ur Physik, Humboldt Universit\"at}
\centerline{\em Invalidenstra\ss{}e 110, 10115 Berlin, Germany}

\vspace{3truecm}

%%%%%%%%%%%%%%%%%%%%%%%%%%%%%%%%%%%%%%%%%%%%%%%%%%%%%%%%

\begin{abstract}

Domain wall solutions have attracted much attention due to their
relevance for brane world scenarios and the holographic RG flow.  In
this talk I discuss the following aspects for these applications: (i)
derivation of the first order flow equations as Bogomol'nyi bound;
(ii) different types of critical points of the superpotential; (iii)
the superpotential needed to localize gravity; (iv) the constraints
imposed by supersymmetry including an example for an $N$=1 flow and
finally (v) sources and exponential trapping of gravity.

\end{abstract}

\vspace{1truecm}

\begin{flushleft}
{\em Talk presented at the RTN-workshop 
"The Quantum Structure of Spacetime and \newline the
Geometric Nature of Fundamental Interactions", Berlin, October 2000
}
\end{flushleft}

%%%%%%%%%%%%%%%%%%%%%%%%%%%%%%%%%%%%%%%%%%%%%%%%%%%%%%%%%%%%%%%%

\newpage

%%%%%%%%%%%%%%%%%%%%%%%%%%%%%%%%%%%%%%%%%%%%%%%%%%%%%%%%%%%%%%%%%

\section{Introduction}
%%%%%%%%%%%%%%%%%%%%%%%%%%%%%%%%%%%%%%%%%%%%%%%

In a spacetime of $d$ dimensions domain walls appear as
$(d-1)$-dimensional objects, that separate the spacetime in two
regions corresponding to different vacua. A well-known example is the
D8-brane solution of massive type IIA supergravity. Like any other
brane solution in supergravity, also domain walls require for
stability a gauge potential.  It is a ($d-1$)-form potential, which
couples naturally to the worldvolume of the domain wall and its
$d$-form field strength is dual to a (negative) cosmological
constant. In other words, a domain wall is a source for a cosmological
constant. In a more general setting with non-trivial couplings to
scalar fields, it is not only a constant, but a potential
$V(\phi)$. For the D8-brane e.g., it is the dilaton potential: $V =
e^{-2 \phi} m^2$, with $m$ as the mass parameter of massive type IIA
supergravity.

There are two kinds of domain walls. For the ``good'' ones the
potential has extrema yielding ``good'' vacua of the
theory. Typically, the extrema are not isolated and have still flat
directions corresponding to remaining moduli (unfixed scalar
values). Unfortunately, many potentials have a run-away behavior as
for the D8-brane and therefore the corresponding scalars do not settle
down at a critical point, but ``run-away''.  These run-away potentials
appear typically in string compactifications, where the run-away
scalar is the dilaton and one refers often to them as ``dilatonic
walls'', for a review of domain walls see \cite{030}.  As consequence,
these ``bad'' domain walls do not allow for a flat space or
anti-deSitter (AdS) vacuum and exhibit a singularity.  Due to the
experimental evidence of a small positive cosmological constant, which
may further decrease in the cosmological evolution, potentials of this
type are widely discussed in cosmology, e.g.\ under the name
quintessence.

We refer to domain walls as kink solutions that interpolates between
two extrema. It may happen that the extrema are $Z_2$-symmetric, but
in general the extrema are different and the domain wall describes the
transition from one vacuum to another one.  In a cosmological setting,
these domain walls may describe a cascade of transitions of false
vacua towards the true vacuum with zero or very small cosmological
constant.  These solutions are the so-called  ``thick
walls''.

As consequence of the AdS/CFT correspondence \cite{250,260,270} domain
walls are expected to encode many information about field
theory. There are especially two application that attracted much
attention recently: (i) the (holographic) supergravity picture for the
renormalization group (RG) flow and (ii) the brane world scenario
yielding an ``alternative to compactification'' by trapping gravity on
the wall (brane) \cite{280,230}.

In this lecture we will summarize different aspects relevant for these
applications.  We start with a discussion of domain walls as solutions
in (super) gravity and derive the first order flow equations of the
scalars. These flow equations are expected to encode the
renormalization group (RG) flow in the dual field theory.  We comment
on the importance of IR attractive critical points of the
superpotential for the brane world scenarios and as an illustrative
example we will consider the Sine-Gordan model. In section 4 we
discuss the constraints imposed by supersymmetry and its implication
for no-go theorems. Many brane world
scenarios start with an $AdS_5$ vacuum and introduce sources to
cut-off ``un-wanted'' pieces of the space. This procedure is justified
if the scalar fields in the asymptotic vacua are fixed by ``correct''
critical points of the superpotential. We will end with a discussion
on this subject.

%%%%%%%%%%%%%%%%%%%%%%%%%%%%%%%%%%%%%%%%%%%%%%%%%%%%%%%%

\section{Scalar flow equations in supergravity}

%%%%%%%%%%%%%%%%%%%%%%%%%%%%%%%%%%%%%%%%%%%%%%%%%%%%%%%%%

We refer to domain walls as kink solutions interpolating between
extrema of a potential $V$ giving different vacua of the theory:
it is a flat spacetime if $V_{extr}=0$, a deSitter vacuum for
$V_{extr}>0$ whereas for $V_{extr}<0$ an anti-deSitter (AdS) vacuum.
The latter case arises naturally in supergravity and implies a
negative vacuum energy.  In the bosonic case, it immediately rises the
question for stability of the vacuum, which is ensured if the
potential is expressed in terms of a superpotential $W$ \cite{070} as
\be020
V = 6 \, \Big(\, {3 \over 4} \, g^{AB} \partial_A W \partial_B W - W^2 \Big)
\ee
and the vacuum is given by an extremum of $W$, i.e.\ $d W
=0$. This also implies an extremum of $V$, but the opposite
is not true. Typically $V$ has more extrema which are rarely stable.
If $W_{extr.}=0$ we obtain a flat space, otherwise an AdS spacetime.
For specific superpotentials, as we will discuss below, these models
can be embedded into supergravity and the scalars saturate the
Breitenlohner/Freedman bound \cite{080}. To be more specific, let us
expand the potential around a given extremum and let us denote the
eigenvalues of the Hessian of the superpotential with $\Delta_{(A)}$,
i.e.\
\be030
\Big( g^{AC} \partial_C \partial_B W \Big)_0 = {1 \over 3} 
	\, \Delta_{(A)} \,   W_0 \, \delta^A_{\ B}
\ee
with $W_0^2 \sim - \Lambda$ as the negative cosmological constant.  On
the other hand the scalar masses are eigenvalues of the mass matrix
$M^A_{\ B} = \Big(\partial^A \partial_B V\Big)_0 = m_{(A)}^2  W_0^2 
\delta^A_{\ B}$, where we again absorbed the mass dimension by the
cosmological constant. Both dimensionless parameters, $\Delta_{(A)}$
and $m_{(A)}$, are related by
\be040
m_{(A)}^2 = \Delta_{(A)} (\Delta_{(A)} - 4) \qquad {\rm or} \qquad
\Delta_{(A)} = 2 \pm \sqrt{4 + m_{(A)}^2} \ .
\ee
So, the Breitenlohner/Freedman bound $m_{(A)}^2 \geq -4$ is ensured
for any real $\Delta_{(A)}$. Note, due to the negative curvature the
lightest scalar fields in AdS spaces have naturally negative
masses. Naively one may argue that this causes naively an instability,
because a mass of a particle can be seen as the minimal energy
necessary to create it out of the vacuum and a negative mass seems to
allow to create an arbitrary number of particles without spending any
energy. This is true for a flat space vacuum, but for an AdS space one
has to be careful in the definition of the energy \cite{090,100} and
this instability does not occur if the Breitenlohner/Freedman bound is
fulfilled.  In the deSitter case the vacuum energy is positive, which
however implies a non-vanishing temperature and deSitter solution
suffer thus from the usual thermodynamical instabilities. As other
solutions with non-vanishing temperature, deSitter solutions break
supersymmetry and we will ignore them furthermore.

Vacua in supergravity are often associated with a {\em negative maximum}
of the potential $V$, but this is in general not the case. Although
extrema of the superpotential yield $V_{extr}<0$ it does not need to
be necessarily a maximum. A maximum of $V$ implies that the Hessian is
negative definite or $m_{(A)} < 0$, which is the case only if $0 <
\Delta_{(A)} < 4$.  Therefore, the potential has a maximum only for
these values of $\Delta_{(A)}$ and especially it has a negative
minimum whenever $\Delta_{(A)} < 0$, which we will discuss below as IR
attractive fixed points.

To be more concrete, we are interested in flat domain walls with a metric
\be050
ds^2 = e^{2U} \Big( -dt^2 + d\vec x^2 \Big) + dy^2 \ .
\ee
This ansatz preserves the Poincar\'e symmetry if the fields depend
only on the transverse coordinate, i.e.\ $U = U(y)$.  For these flat wall
solutions, the gauge fields are trivial and the action reads
\be060
S = \int_M \Big[ {R \over 2} - {1 \over 2} g_{AB} \partial \phi^A 
	\partial \phi^B - V \Big] - \int_{\partial M} K \ .
\ee
We included also a surface term with the outer curvature $K$ and  the
scalar fields $\phi^A = \phi^A(y)$ parameterizing a space ${\cal M}$
with a metric $g_{AB}$. 

The Poincar\'e invariance of the ansatz (\ref{050}) implies that all
worldvolume directions are Abelian isometries, so that we can
integrate them out.  For our ansatz the Ricci scalar takes the form
$R= - 20 (\dot U)^2 - 8 \ddot U$ and after a Wick rotation to an
Euclidean time we find the resulting 1-dimensional action\footnote{In our
notation, dotted quantities always refer to $y$-derivatives.}:
\be070
S \sim \int \, dy \, e^{4U} \Big[ - 6 \, \dot U^2 +
	{1 \over 2} g_{AB} \dot \phi^A \dot \phi^B + V\Big] \ .
\ee
In deriving this expression, the surface term in (\ref{060}) was
canceled by the total derivative term.  The equations of motion of
this action describe trajectories $\phi^A= \phi^A(y)$ of particles in
the target space ${\cal M}$ with the metric $g_{AB}$. This is closely
related to an analogous discussion of black holes \cite{610,600,590}.
As a consequence of the 5-d Einstein equations, these trajectories are
subject to the constraint
\be080
- 6 \, \dot U^2 + {1 \over 2} |\dot \phi^A|^2 - V = 0
\ee
with $|\dot \phi^A|^2 = g_{AB} \, \partial_y \phi^A \partial_y \phi^B$.
In order to derive the Bogomol'nyi bound we can insert the potential
and write the action as
\be090
S \sim  \int \, dy \, e^{4U} \Big[ - 6 \, (\dot U \mp W)^2 +
{1 \over 2} \big| \dot \phi^A  \pm 3 \, \partial^A W \big|^2 \Big]
\mp 3  \int dy {d \over dy}\Big[e^{4U} \, W \Big]
\ee
leading to the BPS equations for the function $U=U(y)$ and $\phi^A =
\phi^A(y)$:
\be100
\dot U = \pm W \qquad , \qquad
\dot \phi^A = \mp 3 \, g^{AB} {\partial W \over \partial \phi^B} \ .
\ee
An analogous derivation of these equations can be found in \cite{591}
and if they are satisfied, the bulk part of the action vanishes and
only the surface term contributes. In the asymptotic $AdS_5$ vacuum
the surface term diverges near the AdS boundary ($U \sim y \rightarrow
\infty$) and after subtracting the divergent vacuum energy one obtains
the expected result that the energy (tension) of the wall is
proportional to the difference of the cosmological constants
(topological charge):
\be102
\sigma \sim \Delta W_0= W_{+\infty} - W_{-\infty} \ .
\ee
If one embeds this model into N=2 supergravity \cite{410, 140} the
first BPS equation becomes equivalent to the variations of the
gravitino and the gaugino/hyperino; the tension is given by the
gravitino charge (mass) and can be obtained by Nesters procedure
\cite{220,090,110}.

Our metric ansatz was motivated by Poincar\'e invariance which is not
spoiled by a reparameterization of the radial coordinate. We have set
$g_{yy}=1$, which is one possibility to fix this residual symmetry,
but we can also use this symmetry to solve the first BPS equation: $W\,
dy = \pm dU$, i.e.\ to take $U$ as the new radial coordinate. In this
coordinate system the metric reads
\be110
ds^2 = e^{2U} \Big(-dt^2 + d\vec x^2 \Big) + {dU^2 \over W^2} \ .
\ee
Repeating the same steps as before we obtain the Bogomol'nyi
equations for the scalars
\be120
- \, \dot \phi^A = g^{AB} \partial_B \log |W|^{3}
	=  g^{AB} \partial_B h 
\ee
which follow from the one-dimensional action
\be130
S \sim \int dy \, \Big[ \, 
|\dot \phi^A |^2 + g^{AB} \partial_A h \partial_B h \Big]
= \int dy \, | \dot \phi^A +  g^{AB} \partial_B h|^2 
+ \ ({\rm surface \ term})
\ee
where $h = 3 \log |W|$. As before, the field equations are subject to
the constraint ${1 \over 2} |\dot \phi^A|^2 - g^{AB} \partial_B h
\partial_B h = 0$ and the surface term yields the central charge.
Supersymmetric vacua are given by extrema of $h$ and the number
and type of such vacua can be determined by using Morse theory where
$h$ is called the height function, see \cite{120}.  To be consistent,
$h$ has to be a ``good'' height function, which means especially that
$h$ and all its derivatives are well-defined on ${\cal M}$ and the
Morse inequalities state that the number of critical points is larger
or equal to the sum over all Betti numbers of ${\cal M}$ \cite{420,630}.

Moreover, the height function $h$ and therefore also the
superpotential are monotonic along the flow.  Namely, multiplying eq.\
(\ref{120}) with $g_{ik} \dot \phi^k$ one obtains 
\be190
- \dot h = - \dot \phi^A \partial_A h =
g_{AB} \dot \phi^A \dot \phi^B \geq 0 
\ee
implying that the height function $h$ is a monotonic decreasing
function towards larger values of $U$.  This is the proposed
supergravity analog of the $c$-theorem \cite{290}, but note it is well
defined as long as $h$ is finite, i.e.\ as long as the superpotential
$W$ does not pass a zero or pole.

%%%%%%%%%%%%%%%%%%%%%%%%%%%%%%%%%%%%%%%%%%%%%%%%%%%%%%%%

\section{RG flow and localization of gravity}

%%%%%%%%%%%%%%%%%%%%%%%%%%%%%%%%%%%%%%%%%%%%%%%%%%%%%%%%

Let us now turn to physical applications that made domain wall
solutions of supergravity so popular. The first application
concerns the renormalization group (RG) flow, see \cite{240} and refs
therein. The other application is the localization of gravity on or
near the wall, which is discussed nowadays as the Randall-Sundrum
scenario \cite{230}, although the basic the idea goes back more than
15 years ago \cite{280}.

%%%%%%%%%%%%%%%%%%%%%%%%%%%%%%%%%%%%%%%%%%%%%%%%%%%%%%

\subsection{Holographic RG flow}

%%%%%%%%%%%%%%%%%%%%%%%%%%%%%%%%%%%%%%%%%%%%%%%%%%%%%%%

The holographic RG flow based on the AdS/CFT correspondence
\cite{250,260,270}, which conjectures that AdS gravity is dual to a
conformal field theory (CFT) and the scalar fields in (super) gravity
corresponds to couplings of perturbation in the field
theory. Originally this conjecture has been made for
$S_5$-compactification of type IIB string theory and the dual field
theory on the worldvolume of a D3-brane resides on the boundary of the
$AdS_5$ space.  This original idea has been extended in a way that any
hypersurface of constant radius should have a field theory dual and
the radial coordinate was identified as the energy scale in the field
theory. In this interpretation, the AdS boundary corresponds to the
UV-limit of the field theory and the radial motion translates into the
RG flow towards the IR.

Let us start with some general remarks about the RG flow in field
theory and repeat some well-known facts. Consider a field theory described
by an action
\be510
S = S[{\cal O}_A, g^B]
\ee
with a set of operators ${\cal O}_A$ with couplings $g^B$.  In
classical field theory these couplings are constant, but due to the
renormalization they become scale dependent $g^B = g^B(\mu)$
(running couplings). This scale dependence is fixed by the 
$\beta$-functions
\be520
\mu {d \over d \mu} g^B = \beta^B(g) 
\ee
which can be derived from the renormalization: $g^B \rightarrow
g^B_{R} - \beta^B \, \log{a\over \mu} + {\cal O}(\log^2{a \over
\mu})$, where $a$ is a cut-off and $\mu$ is the RG-scale. Note, the
couplings $g^B$ are not necessarily gauge couplings, but the couplings
to any perturbations of the Lagrangian and the $\beta$-functions do
not necessarily refer to gauge field $\beta$-functions.

\begin{figure} 
\begin{center}
\includegraphics[width=100mm]{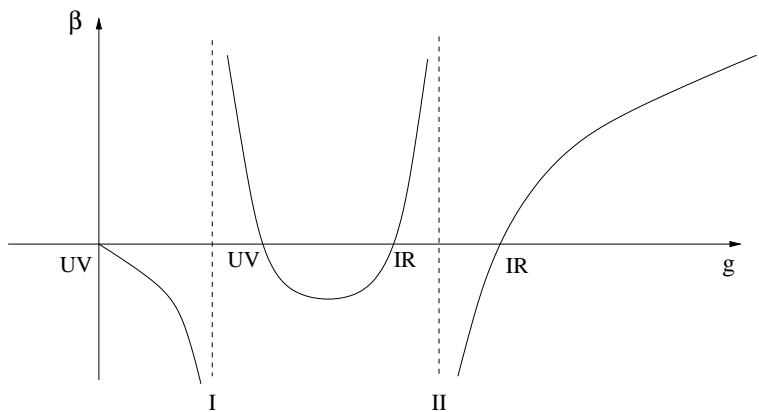}
% \mbox{\epsfig{beta.eps, width=100mm}}
\end{center}
\caption{In general the $\beta$-function may have different zeros,
which are UV- or IR-attractive. In addition, there may be different
poles separating different phases of the field theory.  {From} the
supergravity point of view, the singular line {\bf I} correspond to a
pole in $W$ whereas {\bf II} is a zero in $W$ and in both cases
we expect a singularity in field theory. }
\label{fig2}
\end{figure}

The zeros of the $\beta^B$-functions are especially interesting,
because at these points the theory becomes scale invariant and
therefore finite.  Near this point the operators ${\cal O}_A$ have a
well-defined scaling behavior and the scaling dimensions $\hat
\Delta_A$ are given by the eigenvalues of $\gamma^A_{\ B}$ appearing
in the expansion around a zero $\beta^A(g_0) = 0$
\be530
\beta^A = \gamma^{A}_{\ B} \, \delta g^B + {\cal O}(\delta g^2)
\ee
with $\delta g^A = g^A - g^A_0$.  The equations (\ref{520}) are first
order differential equations describing a flow towards a zero of
$\beta^A$, which are fixed points of the flow. Solving these equations
near a fixed point we find $\delta g^A \sim e^{\hat \Delta^A \log\mu}$
and stability requires that $\hat \Delta^A \log\mu \rightarrow -
\infty$ while we perform scale transformations $\mu \rightarrow
e^{\lambda} \mu$. In an UV scaling: $\lambda \rightarrow + \infty$
this implies $\hat \Delta^A < 0$ whereas in an IR scaling $\lambda
\rightarrow - \infty$: $\hat \Delta > 0$. Hence, we get the picture as
shown in figure \ref{fig2}, that in different scaling limits the
coupling runs to different zeros of the $\beta$-function.  .

Unfortunately in many cases the $\beta$-functions are known only
perturbatively for small couplings and they are often only
well-defined in the UV regime. The IR behavior of many field theories
are out of reach and it would be interesting to have a dual
description which is still valid at points where field theory methods
break down.  Having this motivation, there has been a significant
effort to use the AdS/CFT correspondence to get new information in
field theory.  The translation table is straightforward: recall the
scalar fields $\phi^A$ in gravity correspond to the field theory
couplings $g^A$ and the warp factor in eq.\ (\ref{110}) $e^{U}$
corresponds to the RG parameter $\mu$.  Obviously, the case $U
\rightarrow + \infty$ corresponds to large supergravity length scale
and therefore, due to the AdS/CFT correspondence, describes the UV
regime of the dual field theory. The opposite happens for $U
\rightarrow - \infty$, which is related to small supergravity length
scales and thus encodes the IR behavior of the dual field theory.
Moreover, the $\beta$-function entering the first order differential
equations (\ref{520}) can be translated into the flow equations
(\ref{120}). Extrema of $W$ are fixed points of the scalar flow
equations and translate into fixed points of the RG flow.  As long as
$W \neq 0$ at the extremum, we obtain an AdS vacuum, which is reached
either near the boundary ($U \rightarrow +\infty$) or near the Killing
horizon ($U \rightarrow -\infty$).

\begin{figure} 
\begin{center}
\includegraphics[width=80mm]{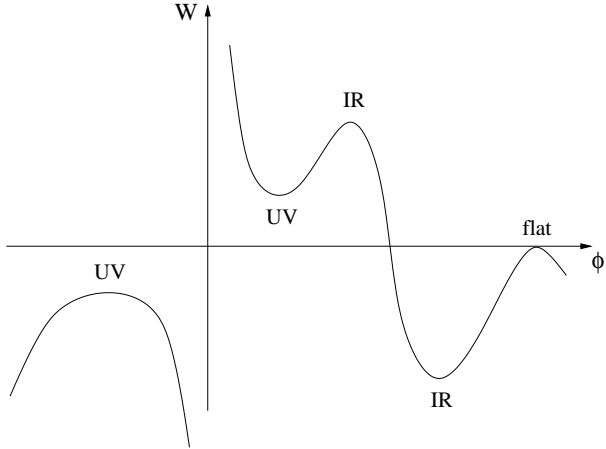}
\end{center}
\caption{An example for a superpotential $W$. At UV extrema
the supergravity solution approaches the boundary of the
AdS space, whereas IR extrema correspond to the Killing horizon.
The supergravity solution is singular at poles of $W$ but
regular at zeros.}
\label{fig1}
\end{figure}

In order to identify the different fixed points we do not need to
solve the equations explicitly; they are determined by the eigenvalues
of the Hessian of the height function $h$.  Let us go back to the BPS
equations and expand these equations around a given fixed point with
$\partial_A W\big|_0 =0$ at $\phi^A = \phi^A_0$. The superpotential
becomes
\be140
W = W_0 + { 1 \over 2} (\partial_A \partial_B W)_0 \, \delta \phi^A \delta
\phi^B \pm \ldots
\ee
with $\delta \phi^A = \phi^A - \phi^A_0$, and the cosmological
constant (inverse AdS radius) is given by $\Lambda = - W_0^2 =
-1/R_{AdS}^2$. Consequently, near the AdS vacuum we find a solution 
of the flow equations (\ref{100}) 
\be180
U = (y-y_0) W_0 \quad , \quad
\delta \phi^A = e^{- {1 \over 3} \, \Delta_{(A)} W_0 (y - y_0)} =
e^{- {1 \over 3}\, \Delta_{(A)} U} 
\ee
with the scaling dimensions introduced in (\ref{030}).  This
approximate solution is valid only if $\delta \phi^A = \phi^A -
\phi^A_0 \rightarrow 0$ in the AdS vacuum with $U \rightarrow \pm
\infty$ and therefore {\it all} eigenvalues $\Delta^{(i)}$ have to
have the same sign: $\Delta_{(A)} >0$ for UV fixed points ($U
\rightarrow + \infty$), or $\Delta_{(A)} <0$ for IR fixed points ($U
\rightarrow -\infty$).  Equivalently, UV fixed points are minima of
the height function $h= \log|W|^3 $ whereas IR fixed points are
maxima, see figure \ref{fig1}. For this conclusion we assumed that the
scalar metric has Euclidean signature and $W_0>0$.  It is important to
notice that in the definition of the scaling dimensions the matrix
$\Omega^A_{\ B}$ has one upper index and one lower index. It is
straightforward to consider also the possibilities $W_0 <0$ and/or
timelike components of the scalar field metric.  Note, the sign
ambiguity in the BPS equations (\ref{100}) interchanges both sides of
the wall, i.e.\ it is related to the parity transformation $y
\leftrightarrow -y$, which also flips the fermionic projector onto the
opposite chirality.

If the eigenvalues $\Delta_{(A)}$ have different signs for different
scalars, this extremum is IR-attractive for some scalars and
UV-attractive for others and, therefore, is not stable (a saddle point
of $h$). A small fluctuation will initiate a further flow towards a
local maximum or minimum for the scalars which enter the
superpotential.  Recall, in our sign convention larger values of the
radial parameter $U$ corresponds to the UV region and are minima of
the height function $h$. If we start with the UV point ($U = +
\infty$) and go towards lower values of $U$, the $c$-theorem states
that $h$ has to increase, either towards an IR fixed point (maximum)
or towards a positive pole in $h$ ($W^2 \rightarrow \infty$), which is
singular in supergravity and corresponds to $c_{CFT} = 0$.  On the
other hand, if we start from an IR fixed point ($U=-\infty$) and go
towards larger values of $U$, due to the $c$-theorem $h$ has to
decrease, either towards a minimum (UV fixed point) or towards a
negative pole ($W^2 \rightarrow 0$), which is {\it not} singular in
supergravity.  An example is the asymptotically flat 3-brane, where
the height function parameterizes the radius of the sphere, which
diverges asymptotically (indicating decompactification) and runs
towards a finite value near the horizon which is IR attractive in our
language.

In summary, there are the following distinct types of supergravity 
flows, which are classified by the type of the extremum of the height
function or superpotential. Depending on the eigenvalues of the
Hessian of $h$, the extrema can be IR attractive (negative
eigenvalues), UV attractive (positive eigenvalues) or flat space
(singular eigenvalues). Generalizing the above discussion and allowing
also possible sign changes in $W$, the following kink solutions are
possible:

$(i)$ flat $\leftrightarrow$ IR

$(ii)$ IR  $\leftrightarrow$ IR

$(iii)$ IR  $\leftrightarrow$ UV

$(iv)$ UV  $\leftrightarrow$ UV \qquad (singular wall)

$(v)$ UV $\leftrightarrow$ singularity \quad ($W^2 = \infty$).

Note, there is no kink solution between a UV fixed point and flat
space, because the $c$-theorem requires a monotonic $h$-function and
the UV point corresponds to a minimum of $h$ whereas the flat space
case is a negative pole. Moreover, if there are two fixed points of
the same type on each side of the wall, $W$ necessarily has to change
its sign implying that the wall is either singular (pole in $W$) or
one has to pass a zero of $W$. In addition, between equal fixed points
no flow is possible (would violate the $c$-theorem) and therefore this
describes a static configuration, where the scalars do not flow. This
is also what we would expect in field theory, where the RG-flows go
always between different fixed points.  Recall, although a zero of $W$
means a singularity in $h$, the domain wall solution can nevertheless
be smooth.  Type $(v)$ walls appear generically for models which can be
embedded into maximal supersymmetric models, for examples see
\cite{500, 520}, whereas models allowing type $(iv)$ walls typically
can not be embedded into maximal supersymmetric
models\footnote{Maximal supersymmetric models typically have only one
UV extremum.}, for an explicit example of this singular flow see
\cite{410}.

Interesting are of course flows towards a confining gauge theory, for
which the Wilson loop shows area law, i.e.\ it scales like
\be374
\langle {1 \over N} {\rm Tr \, P}
e^{\oint A_{\mu} dx^{\mu}} \rangle \sim e^{-c\, A}
\ee
with some constant $c$ and $A$ is the area enclosed by  the Wilson loop.
Writing the 5-d metric as
\be937
ds^2 = f(y) \Big[ -dt^2 + d\vec x^2 \Big] + g(y) dy^2
\ee
the Wilson loop confinement criterion is fulfilled if the warp
factor has a lower bound
\be322
f \rightarrow f_{min} > 0 \quad , \quad g \sim {1 \over y^2} + finite
\quad , \quad {\rm for} \ y \rightarrow 0
\ee
which has been nicely summarized in \cite{320}, but see also
\cite{580} for a review. Notice, the 5-d metric becomes flat for a
flow towards a confining gauge theory.

\begin{figure} 
\begin{center}
\includegraphics[width=100mm]{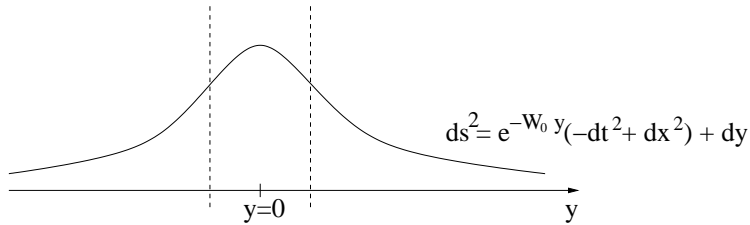}
\end{center}
\caption{In order to localize gravity near the branes
one needs an exponential suppression of the warp factor on both
sides. This means that the scalars have to be fixed by an IR attractive
fixed point of the potential and the superpotential has to change its
sign.}
\label{fig3}
\end{figure}

%%%%%%%%%%%%%%%%%%%%%%%%%%%%%%%%%%%%%%%%%%%%%%%%%%%%%%%%%

\subsection{Localization of gravity}

%%%%%%%%%%%%%%%%%%%%%%%%%%%%%%%%%%%%%%%%%%%%%%%%%%%%%%%%

There is another application that attracted much attention recently,
namely the possibility to localize gravity on a domain wall. This old
idea \cite{280} has recently been discussed as an alternative to
compactification \cite{230}.  The central idea is to employ the
exponential warp factor to suppress any dynamics perpendicular to the
wall, see figure \ref{fig3}. In the language of the RG flow this means
that on both sides of the wall one approaches asymptotically an IR
fixed point, i.e.\ the Hessian of $h \sim \log |W|$ has to be negative
definite at the critical point!  So, we need a potential with two IR
points as shown in figure \ref{fig1}.  Moreover, in order to have an
exponential suppression on both sides, $W$ has to change its sign and it
will have the opposite tension than a (singular) wall separating two
UV critical points.

For the idealized situation, that the space is always AdS (thin wall
approximation), it could be shown that the exponential warp factor
traps the massless graviton mode near the wall \cite{280, 230} and if
the cosmological constant is large enough the fifth direction becomes
invisible -- at least at our low energies. In this picture, our 4-d world
can be seen as a 3-brane embedded in a higher-dimensional space.
However, in the setup that we discussed so far, the scale of the
gravitational and the gauge interaction would be of the same
magnitude, which is not the case in our world. In contrast, in our
low-energy world the gravitational scale is suppressed by many
magnitude and this discrepancy is also known as the gauge hierarchy
problem. In a modified version, the brane world scenario has been
proposed to yield an elegant resolution of this problem
\cite{310,300}. Namely one can introduce a second brane, the so-called
Standard model brane, and brings it close to the IR critical
point. Obviously, the gravitational scale on the Standard model brane
is exponentially suppressed with respect to the former, the so-called
Planck brane. In the spirit of the RG-flow, one may assume that in the
UV regime both brane are close to each other and only in the limit of
low energies our Standard model brane moves closer and closer to the
IR critical point and gravity becomes weaker and weaker.

Let us note, that these solutions necessarily violate the proposed
$c$-theorem (\ref{190}) and there are no-go theorems for constructing
smooth domain walls of this type \cite{440,430}.  This seems to be in
contradiction to the fact that there are smooth solution as in N=1,D=4
supergravity \cite{450}, but also in 5 dimensions as we will see in
the next section. To resolve this puzzle one may not regard these
solutions as flows, but instead as a static configuration where both
AdS vacua coexist and not as a decay of one vacuum into another.

%%%%%%%%%%%%%%%%%%%%%%%%%%%%%%%%%%%%%%%%%%%%%%%%%%%%%%%

\subsection{As an example: Sine-Gordan model}

%%%%%%%%%%%%%%%%%%%%%%%%%%%%%%%%%%%%%%%%%%%%%%%%%%%%%%%%

Unfortunately, supergravity as it comes from compactified string or
M-theory exhibits an abundance of UV critical points but almost no IR
critical points. This is of course related to difficulties with
negative tension branes, which would arise in a thin wall
approximation. But the existence of IR critical points is essential
for the brane world scenarios discussed before, only IR critical
points provide the exponential suppression.

As a illustrative example let us consider the Sine-Gordan model where the
superpotential is given by
\be837
W = a + b \, \cos\theta
\ee
and the kinetic term of the angular scalar field $\theta$ is
normalized as $g_{\theta\theta}=0$. The critical points are at 
$\cos\theta = \pm 1$ and the type 
is related to the derivative of the $\beta$-function calculated
at the different critical points
\be492
- \partial_{\theta} \beta|_{\pm} = \Delta =  3 \partial_{\theta}
\partial_{\theta} \log W \Big|_{\pm} = - { 3\, b \over b \pm a}
=\left\{ \ba{ll} < 0 & {\rm IR} \\ > 0 & {\rm UV} \ea \right. \ .
\ee
For $a>b>0$ we have alternately UV and IR critical points,
for $a = b$ we get a flow towards flat space ($dW=W=0$) and
for $a=0$ there are only IR critical points. For this model the
flow equations (\ref{100}) can be solved explicitly \cite{150}, but
let us discuss only two examples. First, for the flat space flow
($a=b$) one finds \cite{150} 
\be632
ds^2 = \Big(1 + e^{-12 a z} \Big)^{-{2 \over 3}} \, 
\Big(-dt^2 + d\vec x^2 \Big)
+ dz^2 \quad , \quad \cos\theta = - \tanh 6az \ .
\ee
This solution describes now a flow from an IR critical point ($\Delta
=- {3 \over 2}$) at $z \rightarrow - \infty$ and flat space time at $z
\rightarrow +\infty$, where the Wilson loop criteria for confinement
(\ref{322}) is fulfilled with $y = e^{z}$.  The IR point is a minimum
for the potential $V$ and the flat space $V=0$ is approached from
above ($ V'' > 0$), see figure 4. Therefore, near this flat space vacuum the
effective cosmological constant is positive and becomes smaller and
smaller providing an example for the quintessence scenario.

The other solution, that we want to mention is the case $a=0$,
which has only IR critical points ($\Delta = -3$). 
The solution can be written as
\be532
ds^2 = \Big(\cosh 6bz\Big)^{-{2 \over 3}} \, 
\Big(-dt^2 + d\vec x^2 \Big)
+ dz^2 \quad , \quad \cos\theta = - \tanh 6bz 
\ee
and provides an explicit realization of the Randall-Sundrum scenario,
which we introduced in the previous subsection.  Since the scalar
$\theta$ is an angle this solution is periodic and we can identify
Plank branes at $\sin\theta = \pm 1$ and the Standard model branes
near the fixed points $\sin\theta =0$. This interpretation becomes
obvious in the thin wall approximation $b \rightarrow \infty$.

\begin{figure} 
\begin{center}
\includegraphics[width=70mm]{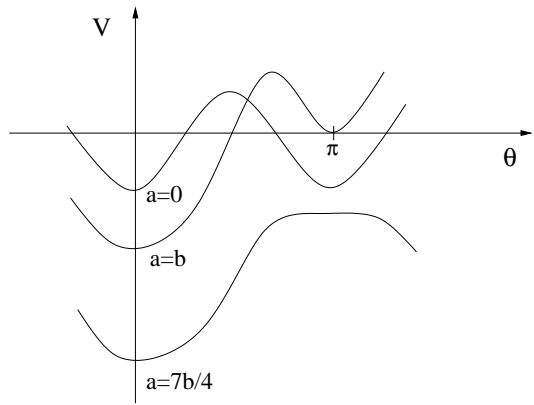}
\end{center}
\caption{This figure shows the potential $V$ for the the superpotential
(\ref{837}). For positive $a,b$ the supersymmetric extrema at $\theta =0$
are always infra-red attractive, whereas the nature of the 
critical point at $\theta = \pi$ depends on the value of $a/b$.
If $4 a = 7b$ one finds $\Delta =4$ and the
corresponding scalar becomes massless.
}
\label{fig4}
\end{figure}

Let us also mention the case, where $\Delta =4$ where the
scalar becomes massless. It happens for $a/b={7 \over 4}$ and 
corresponds to the plateau in figure 4. This is exactly the point, where
the extremum of $V$ converts from a minimum ($\Delta > 4$) into a maxima
($0<\Delta < 4$); see eq.\ (\ref{040}). {From} the field theory point
of view this is an UV critical point whereas $\theta =0$ corresponds
to the IR regime. Notice, whenever $a>b$ the solution has both types
of critical points and when $a<b$ there are only IR critical points
with $\Delta <0$.

The superpotential (\ref{837}) yields a supergravity potential of the
form $V = a_1 + a_2 \cos\theta + a_3 \cos2\theta$, where the
coefficients depend on $a,b$. Potentials of this type are generated
naturally by an instanton/monopole condensation and have an long
history in 4-d gauge theory in the discussion of confinement
\cite{380}, but where also discussed for 5d domain walls, e.g.\
recently in \cite{400}.  Let us concentrate on the discussion of a
single cosine potential and we refer  to \cite{390} for more details.
In 4-d gauge theory this potential can be derived by summing over
instantons and anti instantons in a dilute gas approximation, i.e.\
widely separated non-interacting instanton and anti instantons. In
fact, from the topological term $\int \theta tr(F \wedge F)$ we obtain
after summing over instantons and anti-instantons a cosine potential
for the axion
\be846
\sum_{n, \bar n} {1 \over n!} {1 \over \bar n!} e^{i \theta (n-\bar n)}
= e^{ 2 \cos\theta} \ .
\ee
But can something similar also happen in 5-d supergravity?  The answer
is yes, if we replace the Yang-Mills instantons by M5-brane
instantons. To be more concrete, 5-d supergravity can be obtained from
M-theory compactification and an M5-brane instanton background
translates into a gas of point-like sources in 5 dimensions with $dG =
n \delta^{(5)}$, i.e.\ the 5-branes wrap the 6-d internal space.  The
compactification of the topological Chern-Simons term $\int C\wedge G
\wedge G$ yields than a topological term $\int \theta dG$. So if $dG $
is non-trivial this term is the 5-d analogue of the familiar universal
axionic coupling discussed above and as in 4 dimensions we expect that
the sum over an M5-brane instanton gas will reproduce the cosine
potential.

%%%%%%%%%%%%%%%%%%%%%%%%%%%%%%%%%%%%%%%%%%%%%%%%%%%%%%%%%

\section{Supersymmetry: What is possible and what not?}

%%%%%%%%%%%%%%%%%%%%%%%%%%%%%%%%%%%%%%%%%%%%%%%%%%%%%%%%%

Imposing supersymmetry, the BPS equations (\ref{100}) become the
fermionic supersymmetry variation for the gravitino and
gaugino/hyperino depending on the type of scalars. In addition,
supersymmetry puts severe constraints on the superpotential $W$. If we
have four unbroken supercharges as for N=1,D=4 supergravity, any
holomorphic $W$ is allowed. On the other hand, if we have eight
supercharges, as for N=2 supergravity in four and five dimensions, the
superpotential has to come from a gauging of global
isometries. Restricting to 5-d N=2 supergravity, the scalar fields can
be in three different multiplets: vector-, tensor- or hypermultiplets.
In ungauged supergravity vector- and tensormultiplets are equivalent;
they are dual to each other, which is not the case in gauged sugra,
see \cite{530} for a discussion of non-trivial tensormultiplets.
The scalars parameterize a direct product space
\be472
{\cal M}_{V/T} \times {\cal M}_H
\ee
where ${\cal M}_{V/T}$ is defined by the cubic equation 
\be395
F = {1 \over 6} C_{IJK} X^I X^J X^K =1
\ee
and $I$ counts the number of vector- and tensorfields.  The four
scalars of a hypermultiplet can be combined to a quaternion and the
scalar manifolds ${\cal M}_H$ has to be quaternionic.  In gauged
supergravity, one gauges isometries of these scalar manifolds. One can
show \cite{130,330} that the gauging of isometries of
${\cal M}_{V/T}$ does not influence the flow equations for the scalar,
but introduces only a further constraint.

More interesting is the gauging of isometries of ${\cal M}_H$ which
can be gauged with different vector fields $A^I$ and are given by
Killing vectors $k_I^u$: $q^u \rightarrow q^u + k^u_I \epsilon^I$ and
$dq^u \rightarrow dq^u + k^u_I A^I$, where $q^u$ denote the scalars
in the hypermultiplets. Supersymmetry requires now that the
Killing vectors have to be tri-holomorphic, which means that they
can be expressed by Killing prepotentials
\be624
\Omega^x_{uv} k^v_I = - \nabla_u P^x_I
\ee
where $\Omega^x$ is the triplet of K\"ahler forms characterizing the
quaternionic manifold ($x=1..3$) and $\nabla$ is the covariant
derivative with respect to the $SU(2)$-part of the $SU(2)\times
Sp(2n_H)$ holonomy of the quaternionic space, for more details we
refer to \cite{540, 330, 340}.  The real-valued superpotential 
which enters the flow equations becomes \cite{140}
\be352
W^2 = \sum_{x=1}^3 \Big( P^x_I X^I \Big)^2 
\ee
and it is especially simple if the $SU(2)$-valued Killing prepotential
has only one component, say $P^3_I$. In this case, the Killing
prepotential can be shifted by constants
\be537
P_I^3 \rightarrow P_I^3 + \alpha_I
\ee
which are the analogs of the FI terms known in field theory.
If only these terms are turned on, one has the special case
of a gauged  $SU(2)$-R-symmetry \cite{010}
and the superpotential reads
\be183
W = \alpha_I X^I 
\ee
and it depends on the vector scalars $\phi^A$ via the constraint
(\ref{395}).  Let us consider this special case in more
detail. Critical points of $W$ are given by
\be375
\partial_A W = \alpha_I \partial_A X^I(\phi^A) = 0 
\ee
and since $\partial_A X^I$ are tangent vectors on ${\cal M}_V$,
these are points where the vector $\alpha_I$ is normal
to ${\cal M}_V$ \cite{550, 180}. The normal vector is given by $\partial_I F$
and has therefore be proportional to $\alpha_I$,
which becomes the attractor equation \cite{350}
\be836
{1 \over W_0} \alpha_I = \partial_I F \ .
\ee
The proportionality factor has been fixed by contracting the equation
with $X^I$. In many simple cases these equations can solved
explicitly, as for so-called $STU$-model ($F=STU$) \cite{350,620}
(see \cite{370} for the 4d case). By performing the second derivative
we can also determine the type these extrema. Using the formula
\cite{360, 560}
\be834
\partial_A \partial_B W = \alpha_I \partial_A\partial_B X^I
= {2\over 3} g_{AB} W + {\cal O}(\partial_A W)
\ee
one finds that all scaling dimensions as introduced in (\ref{030}) are
\be184
\Delta_{(A)} = + 2 \ .
\ee
The ``+'' sign indicates that we have an UV fixed point and the
``2'' that they are related to mass deformation in field theory.
But, this means especially that there are {\em no} IR critical points
and therefore the flow has necessarily go towards the 
singularity\footnote{Note, the analog situation in 4 dimensions
may allow for a loophole \cite{570}.}
$F=0$, at which the $\beta$-function has a regular zero \cite{160},
but the spacetime metric is singular.

Therefore, if we have a superpotential depending only on vector
scalars, no smooth flows are possible and especially the lack of IR
critical points means, that brane world scenarios are excluded
\cite{130, 170}.  But notice, this conclusion based on the relation
(\ref{834}) which holds {\em only for vector scalars}. In fact,
turning on hyper scalars and allowing for general Killing
prepotentials $P^x_I = P^x_I(q)$, smooth flows connecting an UV and an
IR attractive fixed point may be possible; see discussion in
\cite{190} where a simple model with one hypermultiplet is considered,
which may be related to the configurations described in \cite{510,
480,490,470,460}.

%%%%%%%%%%%%%%%%%%%%%%%%%%%%%%%%%%%%%%%%%%%%%%%%%%%%%%%

\section{Sources and stability of brane world scenarios}

%%%%%%%%%%%%%%%%%%%%%%%%%%%%%%%%%%%%%%%%%%%%%%%%%%%%%%

It is very typical in supergravity, that solutions are only consistent
if one adds appropriate sources. This is especially necessary for all
(charged) solutions of co-dimensions greater than two, i.e.\  in 10
dimensions all p-brane solutions with $p<7$. All these
solutions are given in terms of a harmonic function $H = H(\vec y)$ on the
transverse coordinates $\vec y$
\be010
\partial^2_{\vec y} H(\vec y) = Q \, \delta(\vec y) \ .
\ee
If the dimension of the transverse $y$-space is greater than two we
need a source for a non-trivial solution, but not necessarily for the
2- or 1-dimensional case. For codimension-2, as the D7-brane, it can
be any holomorphic function whereas for codimension-1 situation, as
for a domain wall, any linear function is harmonic.  On the other
hand, there are good reasons to introduce sources also for
codimension-1 and codimension-2 object.  One reason is that T-duality
should map all D-brane solutions onto one another and this implies
that sources on the rhs are present for all of them.  Another reason
is, that a linear harmonic function has necessarily a zero, which
causes in many cases a curvature singularity. This singularity can be
avoided if we consider the harmonic function: $H = a + b\, |y - y_0|$,
with $a$ and $b$ as some positive constants. The absolute value
obviously corresponds to a source at the position $y=y_0$ and both
sides are $Z_2$ symmetric. Physically, this means, that we cut-off the
singular part and glue together two regular pieces at $y= y_0$.  In
this setup the domain wall can be identified with this singular
source, see \cite{040} for more details. 

For the brane world scenarios there are some subtle points, which one
has to take into account.  To solve the hierarchy problem and for the
localization of gravity it is essential that the warp factor of the
metric has exponential fall-off -- at least asymptotically. This is in
fact the case for the pure $AdS_5$ case, but a cosmological constant
is always a very special limit and in general the low-energy
supergravity has a potential depending on the various scalar
fields. If this potential has an extrema, there exist of course a
solution of the flow equations with constant scalars yielding $dW =0$
and in this case the spacetime is $AdS_5$ everywhere. Next, one may
introduce sources and realize a brane world scenario with an
exponential warp factor. This setup is robust, if we can allow for
small fluctuations of the scalar fields and the exponential
suppression survives -- at least asymptotically. This however is only
the case, if the asymptotic vacua is an IR attractive fixed point with
$\Delta <0$; cp.\ the asymptotic solution (\ref{180}).  If however,
the scalars are fixed at an UV attractive fixed point any small
fluctuation will destroy the exponential warp factor. One still has an
exponential increase in one direction, but in the other direction the
warp factor goes to zero only with a certain power of the radial
distance.  Actually having models as coming from the superpotential
(\ref{183}) without any IR fixed point, any small fluctuations in the
scalars will cast the exponential suppression into a power
suppression. On the other hand having models with an IR fixed point,
the exponential suppression is stable under small fluctuations of the
scalars and in these models we can set the scalars to their critical
value.  Therefore, the challenge in 5-d supergravity is to find
``good'' superpotentials and having this, one may cut-off the regions
close to the UV and IR critical points and continue in a periodic way.
This means that one has to introduce sources and one of them has to
have a negative tension and such objects are not understood.

%%%%%%%%%%%%%%%%%%%%%%%%%%%%%%%%%%%%%%%%%%%%%%%%%%%%%%%%%%%%%%%%

\vskip0.5cm
\noindent
{\large \bf Acknowledgments}

\smallskip
\noindent
I would like to thank the organizers for this nice workshop and the
productive atmosphere and Kathleen Baltzer for interesting comments.
This work is supported by a Heisenberg grant of the DFG.

\newpage

%%%%%%%%%%%%%%%%%%%%%%

%
% ---- Bibliography ----
%
% \nocite{*}                   %this uses *everything* in the .bib file
% \bibliography{proceed00}          %or whatever your .bib file is
% \bibliographystyle{utphys}   %if you use utphys.bst

\providecommand{\href}[2]{#2}\begingroup\raggedright\endgroup

%%%%%%%%%%%%%%%%%%%%%%

\end{document}